%% file: main.tex
\PassOptionsToPackage{table}{xcolor}
\documentclass[sigconf]{acmart}

\AtBeginDocument{}

\setcopyright{acmlicensed}
\copyrightyear{2026}
\acmYear{2026}
\acmDOI{10.1145/XXXXXXX.XXXXXXX}
\acmConference[Internetware '26]{The 17th International Conference on Internetware}{July 18--20, 2026}{Gold Coast, QLD, Australia}
\acmISBN{978-1-4503-XXXX-X/26/07}

\usepackage[T1]{fontenc}
\usepackage[utf8]{inputenc}
\usepackage{microtype}

\usepackage{amsmath}
\usepackage{graphicx}
\usepackage{subcaption}
\usepackage{booktabs}
\usepackage{pifont}
\usepackage{enumitem}
\usepackage{xcolor}
\usepackage[most]{tcolorbox}
\usepackage{tikz}
\usepackage{adjustbox}
\usepackage[normalem]{ulem}

\DeclareRobustCommand{\blackcircle}[1]{%
  \tikz[baseline=(char.base)]{
    \node[circle, fill=black, inner sep=1.2pt] (char)
      {\color{white}\sffamily\footnotesize #1};}}

\usepackage{algorithm}
\usepackage[noend]{algpseudocode}

\newcommand{\diff}[1]{\textsuperscript{\textcolor{blue}{\scriptsize #1$\downarrow$}}}

\definecolor{PhaseColor}{RGB}{170,0,85}
\colorlet{LineNoColor}{black!55}
\definecolor{lightpurple}{RGB}{235,230,250}

\algrenewcommand\algorithmicrequire{\textbf{Require:}}
\algrenewcommand\algorithmicensure{\textbf{Ensure:}}

\makeatletter
\algrenewcommand\alglinenumber[1]{\scriptsize\color{LineNoColor}{#1}}
\makeatother

\newcommand{\Phase}[1]{\State \textcolor{PhaseColor}{\#~#1}}
\algrenewcommand\algorithmiccomment[1]{\hfill\textcolor{PhaseColor}{\#~#1}}

\begin{document}

\title{CodeCoR: An LLM-Based Self-Reflective Multi-Agent Framework for Code Generation}

\author{Ruwei Pan}
\affiliation{%
  \institution{Chongqing University}
  \city{Chongqing}
  \country{China}
}
\email{panruwei@stu.cqu.edu.cn}

\author{Hongyu Zhang}
\authornote{Corresponding author.}
\affiliation{%
  \institution{Chongqing University}
  \city{Chongqing}
  \country{China}
}
\email{hongyujohn@gmail.com}

\author{Chao Liu}
\affiliation{%
  \institution{Chongqing University}
  \city{Chongqing}
  \country{China}
}
\email{liu.chao@cqu.edu.cn}

\input{Abstract}



\maketitle

\section{Introduction}
\input{Introduction}

\section{Related Work}
\input{Related_Work}

\section{Methodology}
\input{Methodology}

\section{Evaluation}
\input{Evaluation}

\section{Discussion}
\input{Discussion}

\section{Threats to Validity}

\input{Threats_to_Validity}

\section{Conclusion}
\input{Conclusion}

\input{Appendix}
\bibliographystyle{ACM-Reference-Format}
\bibliography{references}

\end{document}

%% file: Abstract.tex
\begin{abstract}

Code generation aims to produce code that fulfills requirements written in natural language automatically. Large Language Models (LLMs) like ChatGPT have demonstrated promising effectiveness in this area. Nonetheless, the LLMs often fail to ensure the syntactic and semantic correctness of the generated code. 
Recently, researchers proposed multi-agent frameworks that guide LLMs with different prompts to analyze programming tasks, generate code, and perform testing 
in a sequential workflow. However, the performance of the workflow is not robust as code generation depends on the performance of each agent. 
To address this challenge, we propose CodeCoR, a multi-agent framework that prunes intermediate outputs at different stages to reduce error propagation in sequential code generation workflows.
Specifically, for a given task description, four agents in CodeCoR generate prompts, code, test cases, and repair advice, respectively. 
Each agent generates more than one output and prunes away the low-quality ones. The generated code is tested in the local environment: the code that fails to pass the generated test cases is sent to the repair agent, and the coding agent re-generates the code based on repair advice. 
Finally, the code that passes the highest number of generated test cases is returned to the users.
Our experiments on four widely used datasets, HumanEval, HumanEval-ET, MBPP, and MBPP-ET, demonstrate that CodeCoR outperforms existing baselines (e.g., CodeCoT and MapCoder), achieving an average Pass@1 score of 77.13\%. 

\end{abstract}

%% file: Introduction.tex
Code generation aims to automatically produce code that fulfills the requirements expressed in natural language \citep{david2017program, pan2026toward, pan2026persistent, zhu2025adacoder}. Successful code generation can significantly enhance the productivity and quality of software development, and remains a pivotal area of research in artificial intelligence, natural language processing, and software engineering \citep{gulwani2017program}. In recent years, Large language models (LLMs) such as ChatGPT have demonstrated promising performance in code generation with advanced language understanding and generation capabilities \citep{Zheng2023}. This progress sets the stage for exploring new prompting techniques and frameworks to further improve code generation.

The Chain-of-Thought (CoT) prompting method is commonly utilized to help LLMs better understand tasks by clarifying the prompts through step-by-step reasoning \citep{NEURIPS2022_9d560961}. 
This approach helps an LLM clarify its reasoning process for complex tasks. Building on CoT, researchers have proposed structured prompting techniques.
For example, 
\citet{li2023structured} developed a Structured Chain-of-Thought (SCoT), which mitigates the limitations of CoT by employing a predefined sequence of semantic steps that closely align with programming paradigms, including control structures such as loops and conditional branches. 
\citet{jiang2023self} observed that CoT often suffers from disorganized and inefficient reasoning and introduced a self-planning CoT that guides the prompt generation process based on the code's objectives. 

Recently, researchers proposed multi-agent frameworks that guide LLMs through sequential steps to analyze programming tasks, generate code, and repair bugs.
This sequential workflow is widely adopted in code generation, including ReAct \citep{yao2023reactsynergizingreasoningacting}, Reflexion \citep{shinn2024reflexion}, Self-Edit \citep{zhang2023self}, Self-Planning \citep{jiang2023self}, CodeChain \citep{le2023codechain} and so on.
In the domain of code generation, CodeCoT \citep{huang2023codecot} and  MapCoder \citep{islam2024mapcodermultiagentcodegeneration} are two representative multi-agent frameworks.
CodeCoT \citep{huang2023codecot} involves three agents for task understanding, test case generation, and code generation. The generated code is required to pass the generated test cases; otherwise, the agent will generate new code. Meanwhile, MapCoder \citep{islam2024mapcodermultiagentcodegeneration} is a state-of-the-art multi-agent framework that generates similar step-by-step solution descriptions for a given programming task, then generates the corresponding code, and finally repairs bugs with the test cases provided in testing data. These works demonstrate the effectiveness of sequential multi-agent frameworks in code generation.

\begin{figure}[htbp]
\centering
\includegraphics[width=0.9\columnwidth]{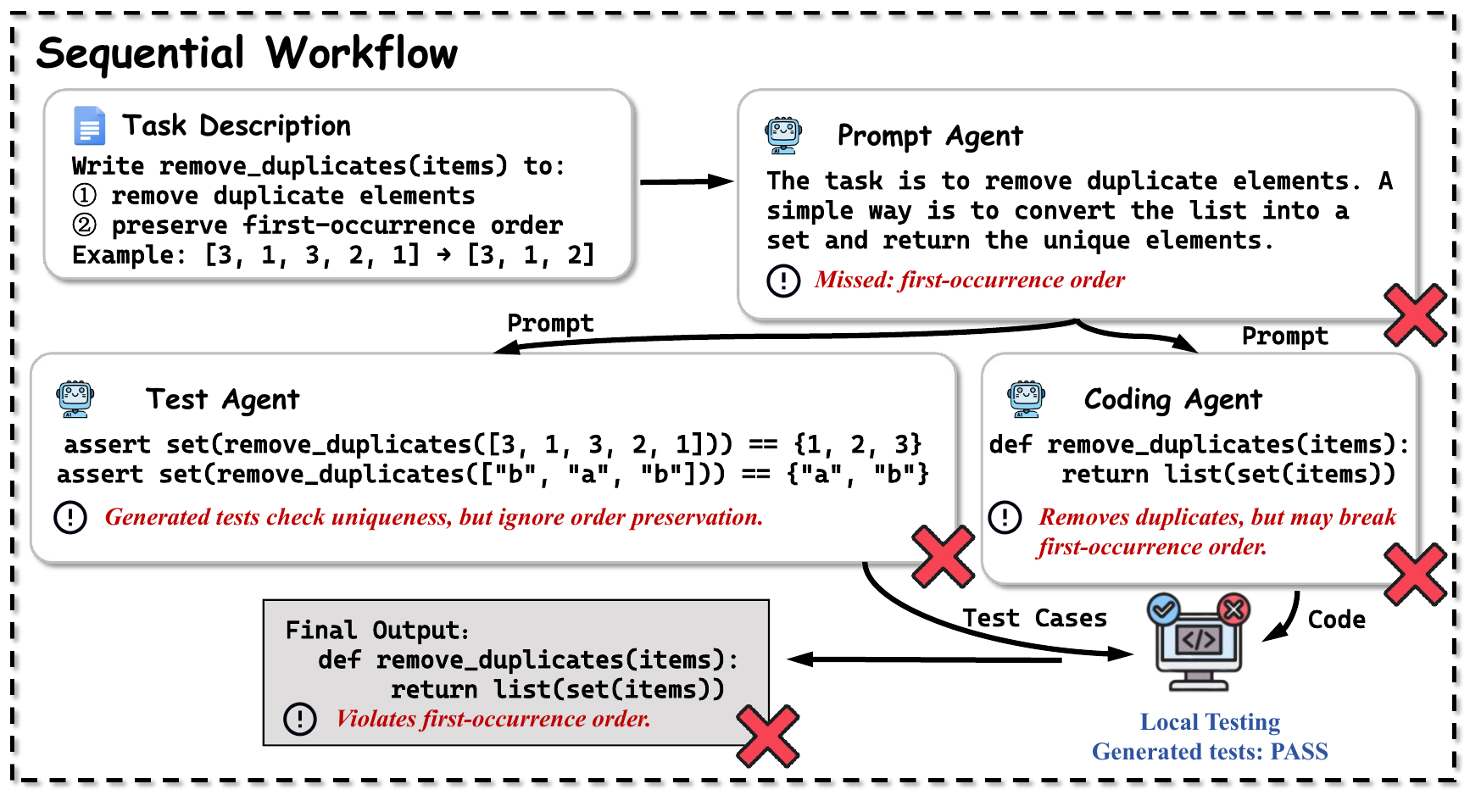}
\caption{An example of misunderstanding in sequential workflow.}
\label{fig:code_misalignment}

\vspace{-10px}
\end{figure}

However, we observed that the performance of existing multi-agent frameworks is not robust, as the performance of each agent 
can significantly affect the quality of the final output. 
As illustrated in Figure \ref{fig:code_misalignment}, in a sequential workflow, once the prompt agent misunderstands the intent of the task description, the misunderstanding will propagate to the follow-up coding agent and test agent. The error is amplified and the generation effort is wasted. 
In Figure~\ref{fig:code_misalignment}, the task requires a function that removes duplicate elements from a list while preserving the order of their first occurrence.
However, the prompt agent focuses on the deduplication requirement but overlooks the order-preservation constraint.
The coding agent then generates code based on \texttt{set()}, which removes duplicates but does not guarantee the original order.
Meanwhile, the test agent produces test cases that check whether duplicates are removed, but fail to verify whether the first-occurrence order is preserved.
As a result, the generated code can pass the generated tests but still violate the original requirement.
This example highlights the need for a more robust approach that can prune low-quality outputs during the code generation process rather than only at the end.

In this study, we introduce CodeCoR (``\textbf{Code} \textbf{Co}llaboration and \textbf{R}eflection"), a self-reflective multi-agent framework for code generation. In CodeCoR, self-reflection is realized by checking and pruning intermediate outputs before they are used by downstream agents, thereby reducing error propagation in sequential multi-agent workflows.
Specifically, CodeCoR consists of four LLM-based agents: \textbf{\textit{1) prompt agent}}, which generates prompts with the CoT technique \citep{NEURIPS2022_9d560961} for task understanding; \textbf{\textit{2) coding agent}}, which generates code for a given programming description; \textbf{\textit{3) test agent}}, which generates test cases according to the task; and \textbf{\textit{4) repair agent}}, which generates repair advice for the code. 
The generated code is tested in the local environment; code that fails to pass the generated test cases is sent to the repair agent, and the coding agent regenerates code based on repair advice. 
Finally, the code that passes the highest number of generated test cases is returned to the users. 
Unlike existing works, 
CodeCoR checks intermediate outputs before they are used by downstream agents. Specifically, it generates multiple candidates and prunes low-quality prompts, test cases, code snippets, and repair advice at their corresponding stages.

We evaluated the proposed framework, CodeCoR, on widely used datasets \citep{chen2021evaluatinglargelanguagemodels,dong2023codescoreevaluatingcodegeneration} including HumanEval, HumanEval-ET, MBPP, and MBPP-ET. 
Experimental results show that CodeCoR outperforms representative prompting and multi-agent baselines, including CodeCoT and MapCoder. In particular, CodeCoR achieves an average Pass@1 score of 77.13\%, compared with 72.8\% achieved by MapCoder.
Moreover, we conducted ablation studies to confirm the necessity of each major component of CodeCoR.

In summary, our major contributions include:

\begin{itemize}

     \item We propose CodeCoR, a self-reflective multi-agent framework involving four collaborative agents (i.e., Prompt Agent, Coding Agent, Test Agent, and Repair Agent) for code generation. The self-reflection in CodeCoR is realized by checking and pruning low-quality or misaligned intermediate outputs before they are passed to downstream agents, reducing error propagation in sequential multi-agent workflows.

    \item We demonstrate the effectiveness of CodeCoR through extensive experiments on multiple datasets. The experimental results show that CodeCoR outperforms representative prompting and multi-agent baselines, and ablation studies further validate the contribution of its major components.

\end{itemize}

%% file: Related_Work.tex
\subsection{Automatic Code Generation}
It has gained increasing attention in software engineering, with early efforts leveraging large language models (LLMs) such as GPT-2 \citep{radford2019language} and GPT-3 \citep{brown2020language} to handle substantial data demands. Subsequent work extended these models to pass given test cases by filtering outputs \citep{chen2021evaluating} and incorporating rule-based sample ranking or complex test generation \citep{li2022competition} \citep{chen2022codet}.

Recent studies increasingly focus on improving code quality through iterative self-revision based on feedback.
Self-Edit \citep{zhang2023self} employs test case results for self-revision, while Self-Correct \citep{welleck2022generating} and CodeRL \citep{le2022coderl} use secondary models to refine output correctness. 
The technique proposed by \citet{olausson2023self} enhances code revision through natural language reflections.
\citet{roziere2021leveraging} and \citet{huang2023scriptoriumws} highlight the importance of large-scale pre-training and robust testing to improve code quality. 
\citet{jiang2023self} propose Self-Planning, which enhances code generation by structuring prompt creation for semantic correctness, but faces challenges with complex tasks, where balancing syntax and semantics remains difficult. This limitation led to the development of the CodeCoR framework.

\subsection{Multi-Agent Coding Models}  
Early multi-agent frameworks focused on basic collaboration strategies \citep{troitzsch2009multi}. \citet{dong2023self} introduced the Self-Collaboration Framework, where LLMs adopt roles like analyst, coder, and tester, using role-specific instructions to enhance task performance. Experiments demonstrate that this self-collaborative approach significantly improves code quality and efficiency. Moreover, SWE-Agent \citep{yang2024swe} and CodeAgent \citep{zhang2024codeagent} improve code correctness through external tools and environment-level interactions.

Recent advancements in multi-agent frameworks include MetaGPT \citep{hong2023metagpt}, which uses human-like SOPs and an executive feedback mechanism to debug and execute code, improving performance on benchmarks like HumanEval. ChatDev \citep{qian2024chatdevcommunicativeagentssoftware} enhances collaboration by allowing agents to seek clarifications, reducing incorrect responses. Additionally, the CodeCoT framework \citep{huang2023codecot} improves code reliability by integrating self-examination to address syntax errors and ensure both semantic and syntactical correctness.
Despite advancements, multi-agent frameworks like MapCoder \citep{islam2024mapcodermultiagentcodegeneration}, which uses retrieval, planning, coding, and debugging agents, face robustness challenges due to their dependency on each agent's performance in a sequential workflow. 
CodeCoR addresses this limitation by checking and pruning intermediate outputs before they are propagated to downstream agents.

%% file: Methodology.tex
\begin{figure*}[htbp]
\centering
\includegraphics[width=0.8\textwidth]{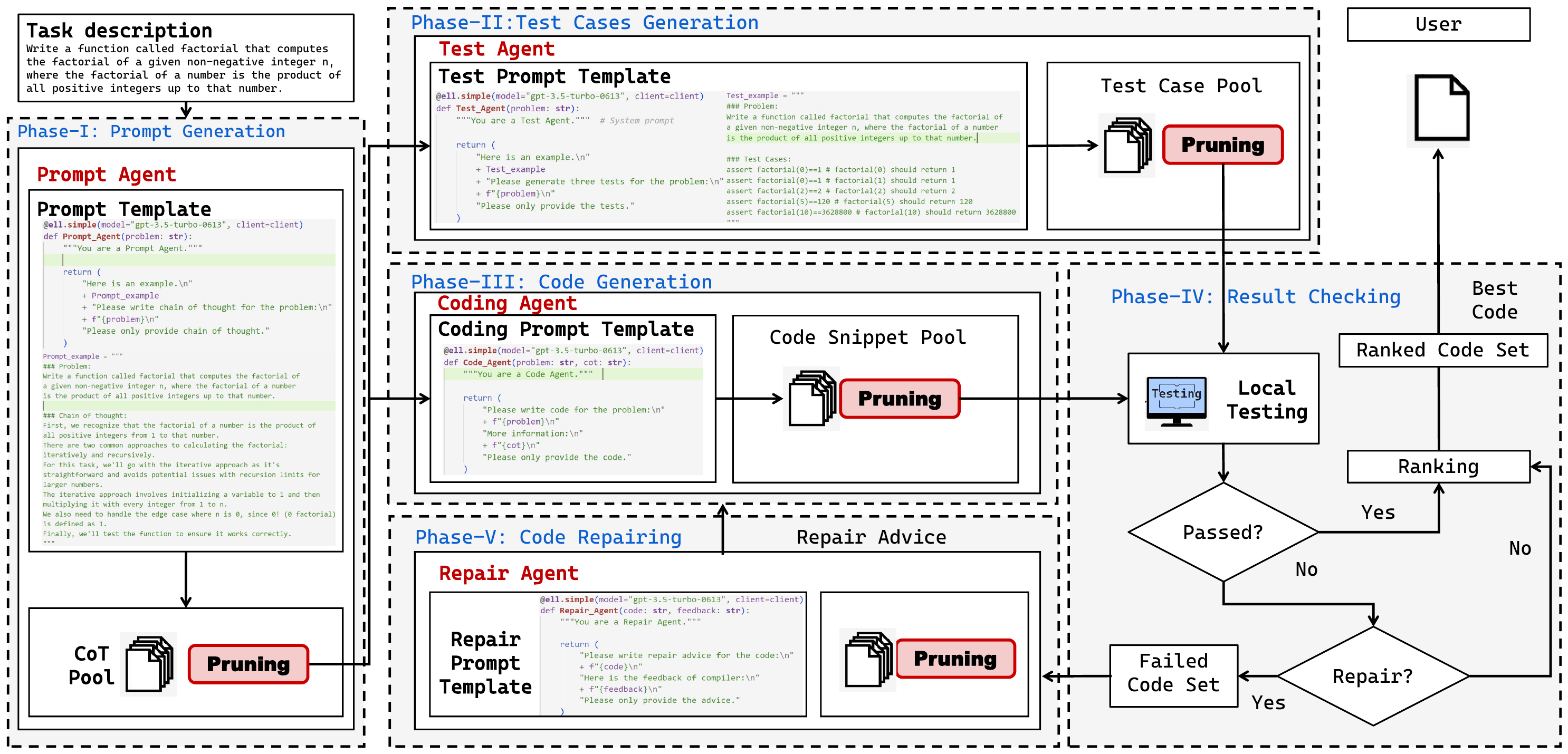}

\caption{Overview of the CodeCoR framework}

\label{fig:overview}
\vspace{-10px}

\end{figure*}

\subsection{Overview}

\begin{figure*}[htbp]
\centering
\includegraphics[width=0.75\textwidth]{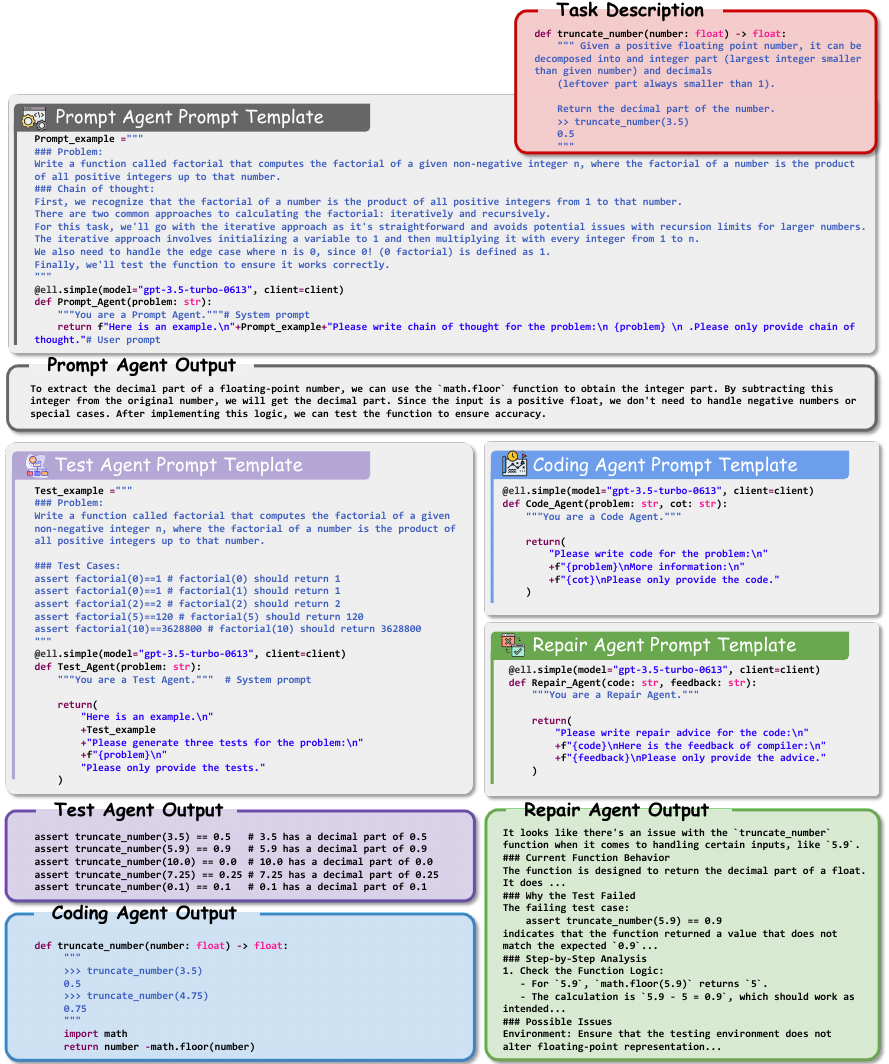}
\caption{Examples of prompts and outputs for four agents.}
\label{fig:prompts}
\vspace{-10px}
\end{figure*}

Figure \ref{fig:overview} depicts the CodeCoR process, which includes four LLM-based agents: \textbf{Prompt Agent, Test Agent, Coding Agent, and Repair Agent}, whose prompts are illustrated in Figure \ref{fig:prompts}. In the Figure, each agent provides an example of its output, when the code task is to write a function that returns the decimal part of a floating-point number. Each agent is assigned a specific task within the generation phases: \textbf{Prompt Generation, Test Case Generation, Code Generation, and Repair}. 

\uline{\textbf{Phase-I: Prompt Generation.}} The process begins with Phase-I: Prompt Generation. Initially, the task description is provided to the Prompt Agent, which generates a series of CoT prompts and stores them in the CoT Pool. The Prompt Agent focuses on generating high-quality CoT prompts.
A pruning method (shown in Section 3.3) is employed to reserve the CoT prompts that decompose the task well, offering detailed step-by-step instructions that facilitate the subsequent generation of code and test cases.

\uline{\textbf{Phase-II: Test Case Generation.}} 
The Test Agent generates a variety of test cases guided by the selected CoT prompts, which are subsequently stored in the Test Case Pool.
To prune low-quality or redundant test cases, the Test Agent uses a pruning method to select only those that effectively assess the executability of the code. This ensures that the test cases are robust and capable of accurately verifying whether the generated code meets the expected requirements.

\uline{\textbf{Phase-III: Code Generation.}} 
Once the CoT prompts are selected, they are passed to the Coding Agent, which generates a variety of code snippets, thereby forming the Code Snippet Pool. The Coding Agent also employs a pruning method to identify and select promising code snippets that demonstrate potential for correctness and efficiency. This approach significantly enhances the quality and accuracy of the final code, reducing the need for further revisions and repairs.

\uline{\textbf{Phase-IV: Result Checking.}} Upon entering Phase-IV, the code generated by the Coding Agent is evaluated against the test cases provided by the Test Agent, and errors are identified by executing the code in a local environment. If all test cases pass, the code is stored in the Ranked Code Set. 
If the code fails some generated test cases, it is added to the Failed Code Set for repair. 
The repair process is controlled by the maximum repair round $R_{\max}$. 
If a code snippet still fails after reaching $R_{\max}$, it is added to the Ranked Code Set.

\uline{\textbf{Phase-V: Repair.}} Otherwise, the process proceeds to Phase-V. 
In this phase, the Repair Agent generates repair advice based on the failed code snippets, and a pruning method is employed to eliminate low-quality advice. 
The advice, along with the erroneous code, is forwarded to the Coding Agent for correction, generating repaired code snippets.
The repaired code snippets undergo further pruning and testing. 
Failed candidates are placed into a new Failed Code Set for the next repair round. 
The repair process stops when the Failed Code Set becomes empty or when the repair round reaches $R_{\max}$. 
Finally, the highest-ranked code is returned to the user.

\begin{figure}[htbp]
\centering
\includegraphics[width=\columnwidth]{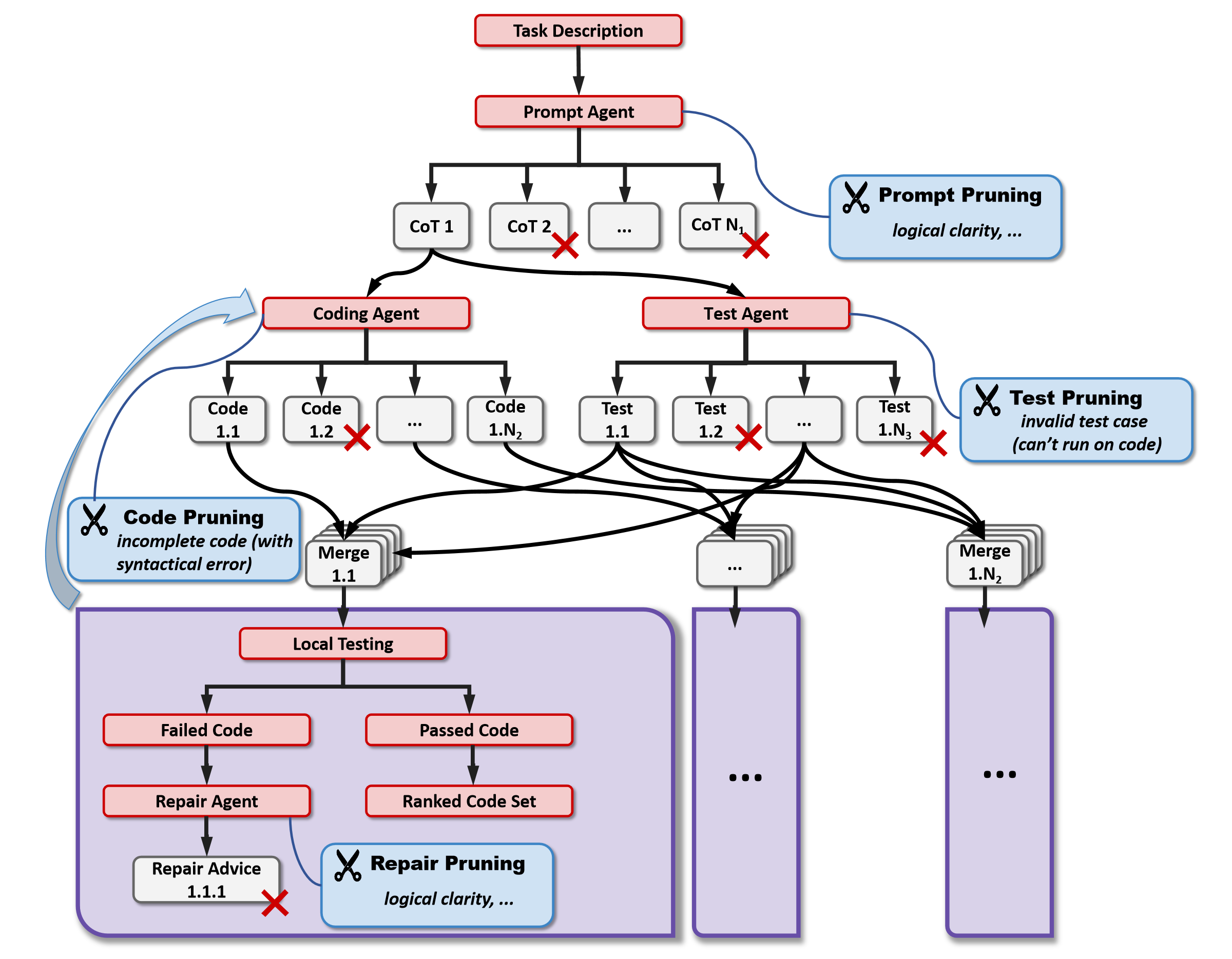}
\caption{The hierarchical structure of the pruning process}
\vspace{-10px}
\label{fig:pruning}
\end{figure}

\subsection{Pruning Methods}

Traditional multi-agent frameworks guide LLMs with different prompts to analyze programming tasks, generate code, test with generated test cases, and repair tested bugs in a sequential workflow. 
However, the workflow is not robust as the code generation depends on the performance of each agent. 
Consequently, CodeCoR realizes self-reflection through pruning methods that check intermediate outputs at different stages before they affect downstream agents, thereby reducing error propagation across the sequential workflow.
Figure \ref{fig:pruning} illustrates the pruning process of the CodeCoR framework, detailing the interactions and functionalities of its components.
Recently, Wang et al.~\citep{Wang_2025} showed that LLMs can be used as auxiliary evaluators for software engineering tasks. 
In CodeCoR, LLM-based evaluation is used only as a lightweight heuristic filter for intermediate outputs, rather than as the final correctness oracle.
Final code correctness is still evaluated using execution-based benchmark tests.
Unlike prior works that rely on fixed metrics or test execution alone, CodeCoR introduces \textbf{pruning prompts} that explicitly guide LLMs to assess the quality of candidate prompts, test cases, and code snippets, allowing the framework to automatically prune low-quality outputs.
By pruning the low-quality outputs, CodeCoR enhances the efficiency of the multi-agent framework. By ranking the results in the Ranked Code Set, CodeCoR outputs the highest-ranked generated code.

\begin{figure*}[htbp]
\centering
\includegraphics[width=0.7\textwidth]{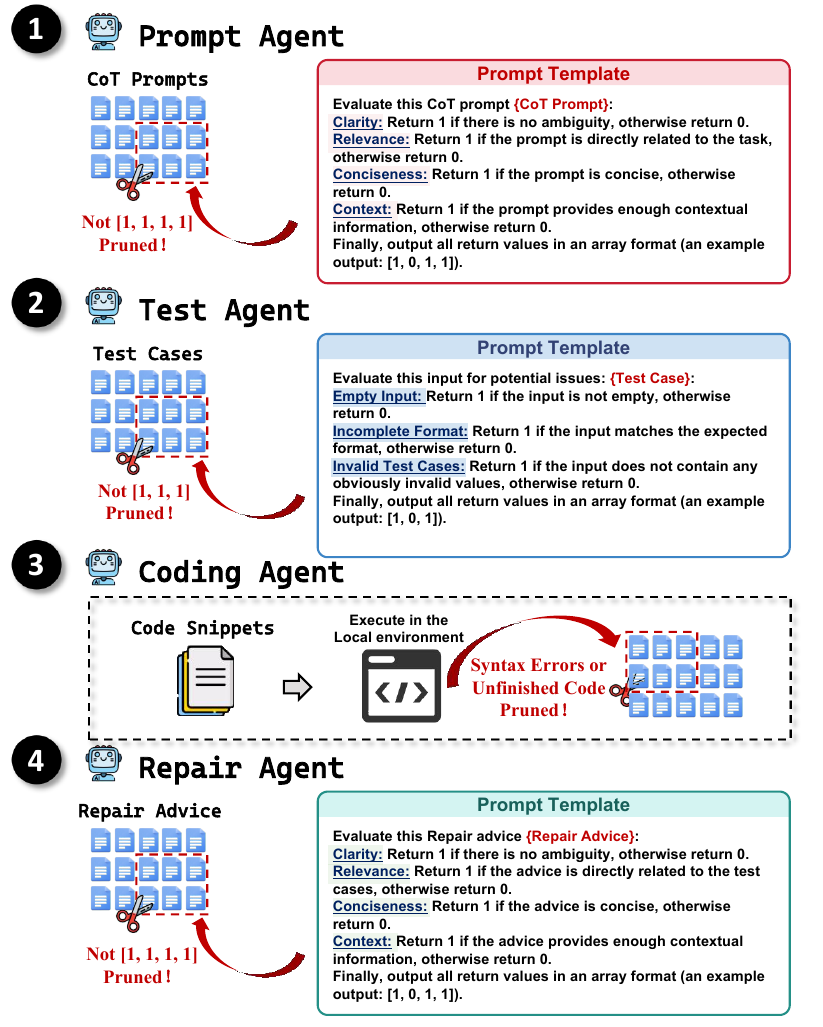}

\caption{The pruning prompts of agents and the workflow for pruning the low-quality outputs}

\label{fig:pruning_method}
\end{figure*}

\uline{\textbf{Prompt Pruning.}} After the Prompt Agent generates a series of CoT prompts for the coding task, it evaluates each prompt’s clarity, relevance, conciseness, and context. 
These criteria are used to check whether a prompt is understandable, task-related, concise, and sufficiently informative. 
An example prompt is shown in Figure \ref{fig:pruning_method}. 
Scores for pruning are assigned by the Prompt Agent based on specific evaluation criteria. 
Each 1 or 0 indicates whether a particular criterion is met or not. The specific criteria are as follows: 
\blackcircle{1} \textbf{Clarity:} whether the prompt or advice is understandable and free from ambiguous instructions. \blackcircle{2} \textbf{Relevance:} whether it focuses on the current task or the observed failure. \blackcircle{3} \textbf{Conciseness:} whether it avoids unnecessary or repetitive content. \blackcircle{4} \textbf{Context:} whether it provides enough information for the downstream agent to continue generation or repair.
We use binary scores because pruning only decides whether a candidate should be passed to the next stage, rather than producing a fine-grained quality ranking.
A prompt is retained only when it receives [1, 1, 1, 1]. 
We use this all-positive rule because failing any criterion may pass ambiguous, irrelevant, redundant, or insufficient information to the next stage. 

\uline{\textbf{Test Pruning.}} After the Test Agent generates test cases, it uses another prompt to classify them. Empty input, incomplete format test cases, or invalid test cases are pruned. Specifically, empty inputs are cases where no data is provided, incomplete format test cases lack significant components necessary for execution, and invalid test cases are cases without the expected data types or falling outside reasonable ranges.

\uline{\textbf{Code Pruning.}} The Coding Agent executes the generated code snippets in the local environment. If code contains syntax errors, such as missing semicolons, unmatched parentheses, unclosed strings, or incorrect indentation, it cannot be compiled and will be pruned by the Coding Agent.

\uline{\textbf{Repair Pruning.}} For each failed code snippet, the Repair Agent provides a single piece of repair advice per repair round. 
If the repair advice cannot meet clarity, relevance, conciseness, and context requirements (as evaluated by the Repair Agent with the specific prompt in \href{https://anonymous.4open.science/r/CodeCoR-3EFC}{Appendix}), the advice is pruned and the failed test cases replace the repair advice. 
Then, the failed test cases are directly submitted, along with the failed code, to the Coding Agent for repair.

By employing these pruning methods, CodeCoR performs self-reflection through intermediate output checking and reduces the propagation of low-quality outputs across stages.

\subsection{Overall Algorithm}

\begin{algorithm}[t]
\caption{CodeCoR Process}
\footnotesize
\begin{algorithmic}[1]
\Require Task description $T_d$, maximum repair rounds $R_{\max}$
\Ensure Final code snippet $C_f$

\State $RankedSet \gets []$
\State $FailedSet \gets []$

\Phase{Phase-I: Prompt Generation (Prompt Agent)}
\State $CoTPrompts \gets generate\_CoT\_prompts(T_d)$
\State $CoTPrompts \gets prune\_or\_fallback(CoTPrompts, prune\_CoT\_prompts)$

\Phase{Phase-II: Test Case Generation (Test Agent)}
\State $TestCases \gets generate\_tests(CoTPrompts)$
\State $TestCases \gets prune\_or\_fallback(TestCases, prune\_tests)$

\Phase{Phase-III: Code Generation (Coding Agent)}
\State $CandidateCodes \gets generate\_code(CoTPrompts)$
\State $CandidateCodes \gets prune\_or\_fallback(CandidateCodes, prune\_code)$

\Phase{Phase-IV: Result Checking}
\For{$codeSnippet \in CandidateCodes$}
    \State $(result, error) \gets execute\_code(codeSnippet, TestCases)$
    \If{$result == \text{"pass"}$} 
        \State $RankedSet.append(codeSnippet)$
    \Else 
        \State $FailedSet.append(codeSnippet)$
    \EndIf
\EndFor

\Phase{Phase-V: Repair (Repair Agent)}
\State $repairRound \gets 0$
\While{$FailedSet \neq \emptyset$ and $repairRound < R_{\max}$}
    \State $repairRound \gets repairRound + 1$
    \State $RepairSuggestions \gets generate\_repair(FailedSet)$
    \State $RepairSuggestions \gets prune\_or\_fallback(RepairSuggestions, prune\_repair)$
    \State $RepairedCodes \gets apply\_repair(RepairSuggestions)$
    \State $NewFailedSet \gets []$
    \For{$codeSnippet \in RepairedCodes$}
        \State $(result, error) \gets execute\_code(codeSnippet, TestCases)$
        \If{$result == \text{"pass"}$} 
            \State $RankedSet.append(codeSnippet)$
        \Else
            \State $NewFailedSet.append(codeSnippet)$
        \EndIf
    \EndFor
    \State $FailedSet \gets NewFailedSet$
\EndWhile

\For{$codeSnippet \in FailedSet$}
    \State $RankedSet.append(codeSnippet)$
\EndFor

\State $C_f \gets select\_best(RankedSet)$
\Return $C_f$

\end{algorithmic}
\end{algorithm}

The CodeCoR process algorithm (Algorithm 1) generates and repairs code to ensure the final output is accurate. 
In \textbf{Phase-I}, based on the task description $T_d$, the Prompt Agent generates a set of CoT prompts \texttt{CoTPrompts} and prunes unpromising ones. 
In \textbf{Phase-II}, the Test Agent uses the selected prompts to generate test cases \texttt{TestCases} and prunes them to ensure that they are robust and executable. 
In \textbf{Phase-III}, the Coding Agent takes \texttt{CoTPrompts} as input to generate multiple candidate code snippets \texttt{CandidateCodes}. It then applies \texttt{prune\_code} to retain promising candidates and improve the efficiency of the subsequent phases. 
For all pruning operations, CodeCoR uses \texttt{prune\_or\_fallback}: if pruning removes all candidates at a stage, the framework keeps the highest-scored candidate according to the corresponding pruning criteria. This prevents the pipeline from stopping due to over-pruning.
In \textbf{Phase-IV}, each snippet in \texttt{CandidateCodes} is tested against \texttt{TestCases} using \texttt{execute\_code}. 
If the execution \texttt{result} is "pass", the snippet is added to the ranked set \texttt{RankedSet}. 
Otherwise, it is added to the failed set \texttt{FailedSet} for repair.
In \textbf{Phase-V}, the Repair Agent generates repair suggestions \texttt{RepairSuggestions} via \texttt{generate\_repair} and prunes them with \texttt{prune\_repair}. 
The suggestions are applied to the failed set \texttt{FailedSet}, producing updated code snippets \texttt{RepairedCodes}. 
These snippets are tested again, and failed snippets are placed into a new failed set for the next repair round. 
The process iterates until \texttt{FailedSet} becomes empty or the repair round reaches $R_{\max}$. 
Any remaining failed snippets after the repair loop are added to \texttt{RankedSet} for final ranking. 
Finally, CodeCoR outputs the highest-ranked code snippet from \texttt{RankedSet} using \texttt{select\_best}, ensuring that the final output $C_f$ is both accurate and efficient.

%% file: Evaluation.tex
\subsection{Research Questions}
This study aims to evaluate CodeCoR by answering the following research questions (RQs):

    \noindent \textbf{RQ1. How effective is CodeCoR for function-level code generation?} 
    \noindent \textbf{RQ2. How effective are the major components of CodeCoR?}

     \noindent \textbf{RQ3. What are the cost implications of  CodeCoR?}
     
     \noindent \textbf{RQ4. How does CodeCoR perform on repository-level code generation? }

\subsection{Datasets}
We evaluate CodeCoR's effectiveness utilizing four widely adopted code generation datasets: HumanEval \citep{chen2021evaluatinglargelanguagemodels} and MBPP \citep{austin2021programsynthesislargelanguage}, along with their enhanced versions, HumanEval-ET and MBPP-ET \citep{dong2023codescoreevaluatingcodegeneration}. HumanEval and HumanEval-ET are designed to provide diverse programming challenges to assess the model's problem-solving capabilities and adaptability. HumanEval is a hand-written evaluation dataset consisting of 164 Python programming problems that assess functional correctness by evaluating language comprehension, reasoning, algorithms, and simple mathematics through unit tests. These problems involve handling specific programming scenarios, necessitating the model to possess comprehension and flexible application skills in Python. 
Meanwhile, MBPP and MBPP-ET offer 
around 1,000 crowd-sourced Python programming problems
aimed at evaluating the model's proficiency in Python syntax and its ability to handle various coding scenarios.

In addition to these function-level datasets, we include DevEval~\citep{li2024deveval} and RAL-Bench~\citep{pan2026ral} for evaluating CodeCoR in more practical software engineering scenarios. DevEval contains repository-level code generation tasks that require models to complete code within existing projects. RAL-Bench further evaluates application-level repository generation, where the generated repositories are assessed using functional system tests. These benchmarks complement HumanEval and MBPP by involving repository context, project-level dependencies, and execution-based validation beyond standalone programming problems.

\subsection{Baselines} 

We show the efficacy of CodeCoR through comparative analyses with a spectrum of prominent LLMs, encompassing both open-source and proprietary variants (see 
\href{https://github.com/Wwstarry/CodeCoR}{our repository} 
for details), including Incoder \citep{fried2022incoder}, CodeGeeX \citep{zheng2023codegeex}, Starcoder \citep{li2023starcoder}, CodeGen-Mono \citep{nijkamp2022codegen}, CodeX \citep{chen2022codet}, GPT-3.5-turbo \citep{openai2024gpt4technicalreport}, ReAct \citep{yao2023reactsynergizingreasoningacting}, Reflexion \citep{shinn2024reflexion}, ToT \citep{yao2023treethoughtsdeliberateproblem}, RAP \citep{hao2023reasoninglanguagemodelplanning}, Self-Edit \citep{zhang2023self}, Self-Planning \citep{jiang2023self}, Self-Debugging \citep{chen2023teaching}, Self-Collaboration \citep{dong2023self}, SCoT \citep{li2023structuredchainofthoughtpromptingcode}, CodeChain \citep{le2023codechain}, INTERVENOR \citep{wang2023intervenor}, CodeCoT \citep{huang2023codecot}, PaLM Coder \citep{chowdhery2023palm}, Claude-instant-1 \citep{luo2023wizardmath}, GPT-4 \citep{openai2024gpt4technicalreport}, MapCoder \citep{islam2024mapcodermultiagentcodegeneration}, CodeTree \citep{li2025codetree}, and DebateCoder \citep{chen2025debatecoder}. 
These methodologies have been empirically demonstrated to yield substantive enhancements in LLM performance across intricate code generation scenarios. 
For function-level baselines, we report results from prior studies when they follow the same benchmark setting~\citep{huang2023codecot,islam2024mapcodermultiagentcodegeneration}.

\subsection{Evaluation Metrics} 
We employ benchmark-specific metrics to evaluate the correctness and quality of the generated code. 
First, we utilize the widely adopted \textbf{Pass@1} metric, which measures the proportion of generated code snippets that correctly perform the intended task on the first attempt without any modifications \citep{chen2021evaluating}. 
Pass@1 is used as the primary metric for function-level benchmarks, since it directly reflects whether the generated code passes the corresponding test cases.
Additionally, to provide a more accurate assessment of the structural and semantic similarity between the generated code and reference implementations, we include \textbf{CodeBLEU}, which enhances token-level evaluation with syntactic features through abstract syntax trees (AST) and semantic features through data flow \citep{ren2020codebleumethodautomaticevaluation}.
We use CodeBLEU only as a complementary metric for code similarity, rather than as a substitute for execution-based functional correctness.
For RAL-Bench, we follow its original evaluation protocol and report \textbf{Fun.}, which measures the average pass rate of functional system tests across generated repositories. Compared with function-level unit tests, Fun. evaluates whether the generated repository satisfies application-level requirements under system-level execution tests.

\subsection{Implementation Details}

For function-level experiments, we use GPT-3.5-turbo 
for CodeCoR and all reproduced baselines, with temperature 0.5 and a maximum output length of 1024 tokens. For DevEval and RAL-Bench, we use GPT-4o-mini with temperature 0.5 and maximum output lengths of 4096 tokens, respectively. CodeCoR uses \(K_p=3\), \(K_t=5\), \(K_c=3\), \(K_r=3\), and \(R_{\max}=3\).
For controlled comparison, we rerun the main baselines with available released code or prompts under the same model, input, decoding setting, and evaluation protocol.
Generated tests are used only for candidate ranking and repair inside CodeCoR. They are not used as the final correctness oracle. All reported Pass@1 results are computed using the official benchmark tests.

\subsection{How effective is CodeCoR for function-level code generation? (RQ1)}
The evaluation results of CodeCoR and the baselines are shown in Table \ref{tab:code_comparison}, which show that CodeCoR achieves state-of-the-art performance compared to baseline models on the HumanEval and MBPP datasets. 
In these evaluations, the GPT-3.5-turbo language model was employed to perform the code generation tasks. 
In Table \ref{tab:code_comparison}, GPT-3.5-turbo achieves the highest average Pass@1 score and shows better performance than other Code LLMs. 
Thus, we adopted GPT-3.5-turbo as the primary backbone model for evaluating our framework, ensuring consistency with prior works and enabling a fairer comparison. In addition, to further assess the generalizability of our approach, we also conducted experiments using more advanced models, including GPT-4-turbo and CodeLlama(34B) shown in Table \ref{tab:combined_humaneval}.
The evaluation results indicate that CodeCoR outperforms prompting-only models on the HumanEval and MBPP datasets.

\begin{table}[h]
    \centering
    \renewcommand{\arraystretch}{1.0}
    \fontsize{7.5pt}{9pt}\selectfont

    \caption{Comparison of code generation models for HumanEval, HumanEval-ET, MBPP, and MBPP-ET datasets (Pass@1 scores). The best approach is highlighted in bold. } 
    \label{tab:code_comparison}

    \begin{adjustbox}{width=0.45\textwidth}

    \rowcolors{2}{lightpurple}{white}

    \begin{tabular}{p{2.0cm}|cccc|c}
        \toprule
        \textbf{Code LLMs} & \textbf{Human} & \textbf{Human} & \textbf{MBPP} & \textbf{MBPP} & \textbf{Avg.} \\
          & \textbf{-Eval} & \textbf{Eval-ET} & \textbf{} & \textbf{-ET} & \\
        \midrule
        Incoder (6.7B)      & 15.2 & 11.6 & 17.6 & 14.3 & 14.7 \\
        CodeGeeX (1.3B)     & 18.9 & 15.2 & 26.9 & 20.4 & 20.4 \\
        Claude-instant-1    & 31.1 & 28.1 & 26.9 & 19.9 & 26.5 \\
        CodeGen-Mono        & 32.9 & 25.0 & 38.6 & 31.6 & 31.5 \\
        PaLM Coder          & 43.9 & 36.6 & 32.3 & 27.2 & 35.0 \\
        StarCoder (15.5B)   & 34.1 & 25.6 & 43.6 & 33.4 & 34.2 \\
        CodeX (175B)        & 47.0 & 31.7 & 58.1 & 38.8 & 43.9 \\
        \textbf{GPT-3.5-turbo} & \textbf{57.3} & \textbf{42.7} & \textbf{52.2} & \textbf{36.8} & \textbf{47.3} \\
        \midrule
        \textbf{Methods}& \textbf{Human} & \textbf{Human} & \textbf{MBPP} & \textbf{MBPP} & \textbf{Avg.} \\
          & \textbf{-Eval} & \textbf{Eval-ET} & \textbf{} & \textbf{-ET} & \\
        \midrule
        Few-Shot           & 67.7 & 54.9 & 65.8 & 48.3 & 59.17 \\
        ReAct              & 56.9 & 49.4 & 67.0 & 45.9 & 54.83 \\
        Reflexion          & 68.1 & 50.6 & 70.0 & 47.4 & 59.03 \\
        ToT                & 54.4 & 42.7 & 65.8 & 40.8 & 50.42 \\
        RAP                & 63.1 & 52.4 & 71.4 & 46.7 & 58.42 \\
        Self-Edit          & 62.2 & 54.3 & 56.4 & 45.9 & 54.70 \\
        Self-Planning      & 65.2 & 48.8 & 58.6 & 41.5 & 53.26 \\
        Self-Debugging     & 61.6 & 45.8 & 60.1 & 52.3 & 54.95 \\
        Self-Collaboration & 74.4 & 56.1 & 68.2 & 49.5 & 62.05 \\
        INTERVENOR         & 75.6 & 54.8 & 69.8 & 47.1 & 61.83 \\
        SCoT               & 60.6 & 53.4 & 67.0 & 51.9 & 58.22 \\
        CodeChain          & 62.8 & 54.3 & 59.1 & 45.5 & 55.68 \\
        Vanilla CodeCoT    & 69.5 & 58.5 & 67.7 & 48.6 & 61.08 \\
        CodeCoT            & 79.3 & 69.5 & 67.7 & 58.1 & 68.90 \\
        MapCoder           & 80.5 & 77.4 & 78.9 & 54.4 & 72.80 \\
        DebateCoder& 67.7 &  65.2 &  63.7  & 37.6   &  58.55   \\
        CodeTree & 85.4    &  79.6   & 61.2   &  43.5 & 67.43 \\
        \textbf{CodeCoR}   & \textbf{86.6} & \textbf{80.5} & \textbf{79.2} & \textbf{65.2} & \textbf{77.13} \\
        \bottomrule
     \end{tabular}
     \end{adjustbox}
\end{table}

On the HumanEval dataset, CodeCoR achieves a Pass@1 score of 86.6\%, whereas Self-Planning \citep{jiang2023self}, SCoT \citep{li2023structuredchainofthoughtpromptingcode}, and Reflexion score 65.2\%, 60.6\%, and 68.1\%, respectively. On the MBPP dataset, CodeCoR scores 79.2\%, higher than Self-Planning, SCoT, and Reflexion \citep{shinn2024reflexion}, which score 58.6\%, 67.0\%, and 70.0\%, respectively. These results indicate that CodeCoR outperforms existing prompting strategies.
When comparing CodeCoR with multi-agent frameworks, CodeCoR also excels. For instance, CodeCoR achieves a Pass@1 score of 86.6\% on the HumanEval dataset, whereas MapCoder \citep{islam2024mapcodermultiagentcodegeneration}  and CodeCoT \citep{huang2023codecot} achieve  80.5\%, and 79.3\%, respectively. On the MBPP dataset, CodeCoR scores 79.2\%, compared to 78.9\% and 67.7\% for MapCoder and CodeCoT, respectively. These results demonstrate that CodeCoR enhances code generation effectiveness compared to other multi-agent frameworks.

\begin{table}[h]
    \centering
    \caption{Comparison of models on the HumanEval and MBPP datasets based on CodeBLEU scores, including the mean values across both datasets. The numbers in parentheses indicate the percentage drop compared to CodeCoR.}
    \label{table:codebleu_results}
    \resizebox{\columnwidth}{!}{%
    \begin{tabular}{l|c|c|c}
        \toprule
        \textbf{Method}       & \textbf{HumanEval} & \textbf{MBPP} & \textbf{Mean} \\
        \midrule
        Self-Planning & 50.25 (\textcolor{blue}{↓3.64\%})  & 41.62 (\textcolor{blue}{↓20.07\%}) & 45.94 (\textcolor{blue}{↓11.84\%}) \\
        SCoT          & 49.68 (\textcolor{blue}{↓4.74\%})  & 34.22 (\textcolor{blue}{↓34.29\%}) & 41.95 (\textcolor{blue}{↓19.51\%}) \\
        CodeChain     & 49.22 (\textcolor{blue}{↓5.60\%})  & 45.89 (\textcolor{blue}{↓11.88\%}) & 47.56 (\textcolor{blue}{↓8.74\%})  \\
        MapCoder      & 49.88 (\textcolor{blue}{↓4.35\%})  & 48.56 (\textcolor{blue}{↓6.74\%})  & 49.22 (\textcolor{blue}{↓5.55\%})  \\
        \rowcolor{cyan!8}
        \textbf{CodeCoR} & \textbf{52.15} & \textbf{52.08} & \textbf{52.12} \\
        \bottomrule
    \end{tabular}%
    }
 
\end{table}

We also compare CodeCoR with baseline models in terms of CodeBLEU scores on the HumanEval and MBPP datasets. We select four models for comparison. The selection of these four models — SCoT \citep{li2023structuredchainofthoughtpromptingcode}, CodeChain \citep{le2023codechain}, Self-Planning \citep{jiang2023self}, and MapCoder \citep{islam2024mapcodermultiagentcodegeneration} — is based on their representativeness and state-of-the-art performance in code generation. Specifically, SCoT, CodeChain, and Self-Planning are chosen for their exceptional semantic capabilities, representing the most advanced semantic baselines, while MapCoder is selected for its state-of-the-art code generation capabilities. 
Specifically, utilizing the reference solutions and the generated code files in the HumanEval and MBPP datasets, we calculated the CodeBLEU between the generated code and the reference code for each task. For each task, we aligned the standard code with the generated code by task id. Subsequently, we computed CodeBLEU for each task and aggregated these scores. The final average CodeBLEU scores were derived by averaging the scores across all tasks.

The comparison results are shown in Table \ref{table:codebleu_results}. On the HumanEval dataset, CodeCoR achieves an average CodeBLEU score of 52.15, outperforming Self-Planning (50.25) and MapCoder (49.88). Although Self-Planning and MapCoder demonstrate comparable results on some individual tasks, CodeCoR's higher average score highlights its superiority in generating semantically accurate and robust code. On the MBPP dataset, CodeCoR again performs exceptionally, achieving an average score of 52.08, closely aligned with its HumanEval performance. When considering the mean CodeBLEU values across both datasets, CodeCoR maintains its leading position, achieving a mean score of 52.12, surpassing all models, including MapCoder (49.22) and CodeChain (47.56). This superior performance reflects that CodeCoR not only produces code that is structurally and semantically closer to the reference solutions but also generalizes effectively across different datasets.

\begin{tcolorbox}[colback=gray!10, fonttitle=\bfseries]
\textbf{\textit{Answer to RQ1:}}  CodeCoR achieves higher performance on four widely used datasets over existing LLM-based code generation models; the code generated by CodeCoR shows better CodeBLEU to the ground-truth code.
\end{tcolorbox}

\subsection{How Effective Are the Major Components of CodeCoR? (RQ2)}

In this section, we analyze how major components of CodeCoR (as illustrated in Figure \ref{fig:overview}) affect its effectiveness. We compare the effectiveness of CodeCoR with the following four variants:

\blackcircle{1} \textbf{w/o Prompt Agent} This variant of CodeCoR operates without the Prompt Agent. The task description replaces the selected CoT prompts in the CoT Pool and is directly passed to the Test Agent and Coding Agent to guide the generation of test cases and code snippets.

\blackcircle{2} \textbf{w/o Test Agent:} This variant is CodeCoR without the Test Agent, which is responsible for validating the syntactic accuracy of the generated code. The goal is to assess whether the absence of the Test Agent leads to an increase in syntax errors. The generated code is executed directly in the local environment. If the code fails, the code and feedback are forwarded to the Repair Agent.
    
\blackcircle{3} \textbf{w/o Repair Agent:} This variant of CodeCoR removes the Repair Agent, which is responsible for correcting errors in the generated code. Its purpose is to verify whether the Repair Agent can enhance the overall quality and reliability of the code by systematically analyzing and fixing these errors. If the generated code cannot pass the test cases generated by the Test Agent, the code and feedback are directly sent to the Coding Agent for repairs.

\blackcircle{4} \textbf{w/o Pruning Method:} This variant is CodeCoR without the Pruning Method, which makes CodeCoR not follow the traditional sequential multi-agent framework. It aims to verify whether the Pruning Method can optimize the generation process by improving the efficiency and quality of agent interactions. This variant cannot prune the outputs of the agents and selects the high-quality outputs of the framework.

The evaluation results are shown in Table \ref{tab:code_comparison_humaneval}. Each major component affects the performance of CodeCoR. 
The lack of Prompt Agent results in a reduction in Pass@1 for each dataset: 77.4\% for the HumanEval dataset and 70.1\% for the HumanEval-ET dataset. This highlights the importance of the Prompt Agent in providing clear context and correct direction for the task, which plays a key role in maintaining the quality and accuracy of the generated code.

Specifically, the absence of the Test Agent results in significant Pass@1 reductions across various datasets: 45.1\% on HumanEval, 43.3\% on HumanEval-ET, 52.1\% on MBPP, and 40.6\% on MBPP-ET. This shows the importance of syntactic validation in maintaining the quality and reliability of the generated code. 
When the Repair Agent is omitted, there are notable declines in Pass@1 scores: 75.6\% on HumanEval, 75.6\% on HumanEval-ET, 67.7\% on MBPP, and 48.6\% on MBPP-ET. The Repair Agent plays a critical role in self-correction, enabling the system to address and rectify errors effectively. Its absence leads to a significant decrease in the overall quality of the code, as the system becomes less capable of correcting errors autonomously.
When the Pruning Method is removed, Pass@1 scores drop to 77.4\% on HumanEval, 79.2\% on HumanEval-ET, 67.7\% on MBPP, and 58.1\% on MBPP-ET. This module is essential for managing the flow and processing of inputs, ensuring that the generated code remains coherent and of high quality. Its absence can lead to increased noise and errors in the code generation process, resulting in performance degradation.

\begin{table}[H] 
\centering
\caption{The impact of major components of CodeCoR on HumanEval and MBPP datasets (Pass@1)}
\label{tab:code_comparison_humaneval}
\resizebox{\columnwidth}{!}{
\begin{tabular}{l|cccc}
    \toprule
    \textbf{Variant} & \textbf{HumanEval} & \textbf{HumanEval-ET}   & \textbf{MBPP} & \textbf{MBPP-ET}\\
    \hline
    w/o Prompt Agent 
      & 77.4 {\diff{9.2}} 
      & 70.1 {\diff{10.4}} 
      & 58.4 {\diff{20.8}} 
      & 46.4 {\diff{18.8}} \\
    w/o Test Agent 
      & 45.1 {\diff{41.5}} 
      & 43.3 {\diff{37.2}} 
      & 52.1 {\diff{27.1}} 
      & 40.6 {\diff{24.6}} \\
    w/o Repair Agent 
      & 75.6 {\diff{11.0}} 
      & 75.6 {\diff{4.9}} 
      & 67.7 {\diff{11.5}} 
      & 48.6 {\diff{16.6}} \\
    w/o Pruning Method 
      & 77.4 {\diff{9.2}} 
      & 79.2 {\diff{1.3}} 
      & 67.7 {\diff{11.5}} 
      & 58.1 {\diff{7.1}} \\
    \rowcolor{cyan!5} 
    \textbf{CodeCoR} & \textbf{86.6} & \textbf{80.5} & \textbf{79.2} & \textbf{65.2}  \\
    \bottomrule
\end{tabular}
}
\end{table}

\begin{tcolorbox}[colback=gray!10, fonttitle=\bfseries]
\textbf{\textit{Answer to RQ2:}}  Our CodeCoR framework exhibits better performance than its variants, confirming the effectiveness and necessity of its major components.
\end{tcolorbox}

\section{What are the cost implications of CodeCoR? (RQ3)}

The cost implications of multi-agent frameworks are usually higher than those of simpler prompting frameworks. 
Therefore, we evaluate the system-level cost of CodeCoR and compare it with representative baselines. 
Table~\ref{tab:performance_comparison} provides an empirical assessment of different code generation frameworks, including MapCoder, CodeTree, DebateCoder, SCoT, Self-Planning, CodeChain, and CodeCoR. 
The experiment was conducted in the Python environment on the full HumanEval benchmark.
We employed \texttt{psutil}~\citep{rodola2020psutil}, a Python library, for monitoring costs. 
We recorded execution time, CPU usage, disk write, and network I/O to evaluate the system-level cost of CodeCoR. 
The experiments were conducted on a dedicated server to ensure that other services or processes did not influence our measurements.

\begin{table}[htbp]
\centering
\caption{System-level cost comparison on the full HumanEval benchmark.}
\label{tab:performance_comparison}
\resizebox{\columnwidth}{!}{
\begin{tabular}{lccccc}
\toprule
\textbf{Method} 
& \textbf{Run Time(s)} 
& \textbf{CPU Usage(\%)} 
& \textbf{Disk Write(MB)} 
& \textbf{Net Send(MB)} 
& \textbf{Net Receive(MB)} \\
\midrule
MapCoder      & 192.59 & 0.06 & 2.36 & 52.25 & 38.78 \\
CodeTree      & 101.37 & 0.05 & 1.30 & 2.27  & 12.57 \\
DebateCoder   & 255.32 & 0.05 & 3.36 & 17.83 & 27.43 \\
SCoT          & 79.69  & 0.07 & 1.27 & 3.56  & 3.57  \\
Self-Planning & 52.48  & 0.05 & 0.95 & 0.74  & 0.78  \\
CodeChain     & 119.13 & 0.05 & 2.48 & 19.61 & 22.20 \\
CodeCoR       & 91.61 & 0.06 & 0.83 & 22.37 & 39.59 \\
\bottomrule
\end{tabular}
}
\end{table}
 
Table~\ref{tab:performance_comparison} reports the system-level cost comparison on the full HumanEval benchmark.
CodeCoR completes the benchmark in 91.61 seconds, which is faster than MapCoder, CodeTree, DebateCoder, and CodeChain, but slower than Self-Planning and SCoT.
This result shows that CodeCoR does not achieve the lowest runtime among all methods, but it remains efficient among multi-agent and structured code generation frameworks.
For CPU usage and disk writing, CodeCoR remains comparable to or better than most baselines.
Its CPU usage is 0.06\%, close to the other methods, and its disk write volume is 0.83 MB, which is the lowest among the compared methods.
For network traffic, CodeCoR has moderate Net Send but relatively high Net Receive.
This is mainly because CodeCoR involves multiple agents and receives more intermediate information during candidate generation, pruning, and repair.
Although CodeCoR introduces additional network communication due to its multi-agent workflow, its runtime, CPU usage, and disk overhead remain manageable.

\begin{tcolorbox}[colback=gray!10, fonttitle=\bfseries]
\textbf{\textit{Answer to RQ3:}}  
In summary, we evaluated the system-level cost of CodeCoR on the full HumanEval benchmark. 
The results show that CodeCoR has manageable overhead: it achieves competitive runtime, comparable CPU usage, and low disk write cost, while its main extra cost comes from network communication due to the multi-agent workflow.

\end{tcolorbox}

\section{How does CodeCoR perform on repository-level code generation? (RQ4)}

To examine whether CodeCoR can generalize beyond standalone function-level tasks, we evaluate it on two more realistic benchmarks: DevEval~\citep{li2024deveval} and RAL-Bench~\citep{pan2026ral}. 
DevEval focuses on repository-level code generation within existing project contexts, while RAL-Bench evaluates application-level repository generation through functional system tests. 
We compare CodeCoR with seven representative baselines, including CoT, CodeCoT, SCoT, Self-Planning, MapCoder, CodeTree, and DebateCoder under the same backbone model and execution protocol. 
We report Pass@1 for DevEval and Fun. for RAL-Bench, where Fun. measures the average pass rate of functional system tests.

\begin{table}[htbp]
\centering
\caption{Performance comparison on DevEval and RAL-Bench.}

\label{tab:rq4_realistic}
\setlength{\tabcolsep}{4pt}
\renewcommand{\arraystretch}{1.10}
\resizebox{\columnwidth}{!}{
\begin{tabular}{lcc lcc}
\toprule
\textbf{Method} & \textbf{DevEval} & \textbf{RAL-Bench} 
& \textbf{Method} & \textbf{DevEval} & \textbf{RAL-Bench} \\
\cmidrule(lr){2-2} \cmidrule(lr){3-3}
\cmidrule(lr){5-5} \cmidrule(lr){6-6}
& \textbf{Pass@1} & \textbf{Fun.}
& & \textbf{Pass@1} & \textbf{Fun.} \\
\midrule
CoT           & 4.77 & 16.47 & CodeCoT       & 5.70 & 18.85 \\
SCoT          & 2.96 & 16.71 & Self-Planning & 5.15 & 18.14 \\
MapCoder      & 3.07 & 18.38 & CodeTree      & 4.27 & 16.95 \\
DebateCoder   & 4.11 & 17.90 & CodeCoR       & \textbf{8.88} & \textbf{20.05} \\
\bottomrule
\end{tabular}
}

\end{table}

Table~\ref{tab:rq4_realistic} reports the results on DevEval and RAL-Bench.
CodeCoR achieves the best performance on both benchmarks, with 8.88\% Pass@1 on DevEval and 20.05\% Fun. on RAL-Bench.
Compared with the strongest baseline, CodeCoR improves by 3.18 percentage points on DevEval and 1.20 percentage points on RAL-Bench.
These results provide preliminary evidence that CodeCoR can generalize beyond standalone function-level tasks to repository-level and application-level code generation.
The gains suggest that checking and pruning intermediate outputs before downstream propagation can still be useful in more realistic software development settings.

\begin{tcolorbox}[colback=gray!10, fonttitle=\bfseries]
\textbf{\textit{Answer to RQ4:}}  
CodeCoR achieves the best results on both DevEval and RAL-Bench, suggesting its potential for repository-level and application-level code generation.

\end{tcolorbox}

%% file: Discussion.tex
\subsection{Can CodeCoR Work with Other LLMs?}

This section evaluates and assesses the performance of various methods applied to two powerful LLMs: CodeLlama \citep{rozière2024codellamaopenfoundation} and GPT-4 \citep{openai2024gpt4technicalreport}. CodeLlama is a specialized model designed for coding tasks, equipped with advanced training techniques such as infilling and long context handling. Conversely, GPT-4 is a multi-modal model that excels in text comprehension and demonstrates exceptional prowess in complex reasoning tasks.

\begin{table}[ht]
\centering
\caption{Comparison of different methods on HumanEval and HumanEval-ET datasets}
\label{tab:combined_humaneval}
\resizebox{\columnwidth}{!}{ 
\begin{tabular}{l|cc|cc}
\toprule
\textbf{Method}& \multicolumn{2}{c|}{\textbf{GPT-4}} & \multicolumn{2}{c}{\textbf{CodeLlama (34B)}} \\
\cmidrule{2-5}
 & \textbf{HumanEval} & \textbf{HumanEval-ET} & \textbf{HumanEval} & \textbf{HumanEval-ET} \\
\hline
CodeChain & 89.0\textsuperscript{\textcolor{blue}{5.5$\downarrow$}} & 61.6\textsuperscript{\textcolor{blue}{21.9$\downarrow$}} & 
15.9\textsuperscript{\textcolor{blue}{28.0$\downarrow$}} & 14.0\textsuperscript{\textcolor{blue}{23.8$\downarrow$}} \\

SCoT & 78.9\textsuperscript{\textcolor{blue}{15.6$\downarrow$}} & 69.5\textsuperscript{\textcolor{blue}{14.0$\downarrow$}} & 
17.4\textsuperscript{\textcolor{blue}{26.5$\downarrow$}} & 14.9\textsuperscript{\textcolor{blue}{22.9$\downarrow$}} \\

Self-Planning & 83.5\textsuperscript{\textcolor{blue}{11.0$\downarrow$}} & 76.8\textsuperscript{\textcolor{blue}{6.7$\downarrow$}} & 
22.6\textsuperscript{\textcolor{blue}{21.3$\downarrow$}} & 20.1\textsuperscript{\textcolor{blue}{17.7$\downarrow$}} \\

CodeCoT & 86.6\textsuperscript{\textcolor{blue}{7.9$\downarrow$}} & 77.4\textsuperscript{\textcolor{blue}{6.1$\downarrow$}} & 
34.1\textsuperscript{\textcolor{blue}{9.8$\downarrow$}} & 29.9\textsuperscript{\textcolor{blue}{7.9$\downarrow$}} \\

ChatDev & 84.1\textsuperscript{\textcolor{blue}{10.4$\downarrow$}} & 72.7\textsuperscript{\textcolor{blue}{10.8$\downarrow$}} & 
23.6\textsuperscript{\textcolor{blue}{20.3$\downarrow$}} & 20.6\textsuperscript{\textcolor{blue}{17.2$\downarrow$}} \\

MetaGPT & 85.9\textsuperscript{\textcolor{blue}{8.6$\downarrow$}} & 74.0\textsuperscript{\textcolor{blue}{9.5$\downarrow$}} & 
26.5\textsuperscript{\textcolor{blue}{17.4$\downarrow$}} & 23.1\textsuperscript{\textcolor{blue}{14.7$\downarrow$}} \\

MapCoder & 93.9\textsuperscript{\textcolor{blue}{0.6$\downarrow$}} & 82.9\textsuperscript{\textcolor{blue}{0.6$\downarrow$}} & 
42.7\textsuperscript{\textcolor{blue}{1.2$\downarrow$}} & 37.0\textsuperscript{\textcolor{blue}{0.8$\downarrow$}} \\

\rowcolor{cyan!8}
CodeCoR & \textbf{94.5} & \textbf{83.5} & \textbf{43.9} & \textbf{37.8} \\
\bottomrule
\end{tabular}
}
\end{table}

Table \ref{tab:combined_humaneval} compares the Pass@1 scores of different frameworks evaluated on the CodeLlama model.
For instance, in the HumanEval dataset, CodeCoR achieves a score of 43.9\%, which surpasses CodeCoT's 34.1\%, Self-Planning's 22.6\%, SCoT's 17.4\%, and CodeChain's 15.9\%. Similarly, in the HumanEval-ET dataset, CodeCoR scores 37.8\%, outperforming all other prompting methods. 

To demonstrate the practical applicability of CodeCoR, we tested the framework on various datasets including HumanEval and HumanEval-ET using GPT-4. The results indicated significant improvements in accuracy compared to existing methods. For instance, on the HumanEval dataset, CodeCoR achieved a Pass@1 accuracy of 94.5\%, and on the HumanEval-ET dataset, it achieved 83.5\%. These results are significantly higher compared to other methods such as CodeChain, SCoT, Self-Planning, and CodeCoT, as shown in Table \ref{tab:combined_humaneval}. We report additional results on more models and benchmarks in \href{https://github.com/Wwstarry/CodeCoR}
{\textbf{our repository}} for reproducibility.

\subsection{How Many Tokens Are Generated by LLMs for Different Prompt Techniques?}

We performed an experiment to compare the average token usage and cost per problem across different code generation methods on HumanEval and MBPP in Table \ref{humaneval} and Table \ref{mbpp}. As shown in the table below, CodeCoR achieves strong performance with a reasonable cost, maintaining comparable or lower input/output token usage than other multi-agent baselines such as CodeCoT and MapCoder.

While CodeCoR incurs slightly higher cost than some single-agent methods, it achieves significantly better performance and reduces cost compared to other multi-agent frameworks, demonstrating a better balance between effectiveness and efficiency.

\begin{table*}[h!]
    \centering
    \begin{minipage}{0.48\linewidth}
        \centering
        \caption{Comparison of Cost and Token Usage for HumanEval Dataset}
        \vspace{-7px}
        \label{humaneval}
        \resizebox{0.8\columnwidth}{!}{%
        \begin{tabular}{lccc}
            \toprule
            \textbf{Methods} & \textbf{Cost (\$)} & \textbf{In-token} & \textbf{Out-Token} \\
            \midrule
            Zero-shot     & 0.0028 & 119.1 & 357.6 \\
            Few-shot      & 0.0016 & 374.6 & 58.0  \\
            CoT           & 0.0047 & 504.3 & 293.9 \\
            Self-planning & 0.0062 & 521.6 & 364.1 \\
            SCoT          & 0.0058 & 508.8 & 384.2 \\
            CodeCoT       & 0.0071 & 535.5 & 428.7 \\
            MapCoder      & 0.0083 & 398.5 & 562.0 \\
            \rowcolor{cyan!8}
            \textbf{CodeCoR} & \textbf{0.0064} & \textbf{517.9} & \textbf{403.2} \\
            \bottomrule
        \end{tabular}}
    \end{minipage}
    \hfill
    \begin{minipage}{0.48\linewidth}
        \centering
        \caption{Comparison of Cost and Token Usage for MBPP Dataset}
       \vspace{-7px}
        \label{mbpp}
        \resizebox{0.8\columnwidth}{!}{%
        \begin{tabular}{lccc}
            \toprule
            \textbf{Methods} & \textbf{Cost (\$)} & \textbf{In-token} & \textbf{Out-Token} \\
            \midrule
            Zero-shot     & 0.0025 & 45.1  & 340.5 \\
            Few-shot      & 0.0022 & 578.8 & 157.0 \\
            CoT           & 0.0044 & 510.4 & 390.5 \\
            Self-planning & 0.0059 & 578.9 & 488.1 \\
            SCoT          & 0.0052 & 559.7 & 479.6 \\
            CodeCoT       & 0.0050 & 534.0 & 449.1 \\
            MapCoder      & 0.0069 & 580.3 & 383.1 \\
            \rowcolor{cyan!8}
            \textbf{CodeCoR} & \textbf{0.0047} & \textbf{521.2} & \textbf{423.6} \\
            \bottomrule
        \end{tabular}}
    \end{minipage}
    \vspace{-8px}
\end{table*}

\subsection{How does the number of repair rounds affect the performance of the agents?}

\begin{figure}[ht]
    \centering
    \includegraphics[width=0.28\textwidth]{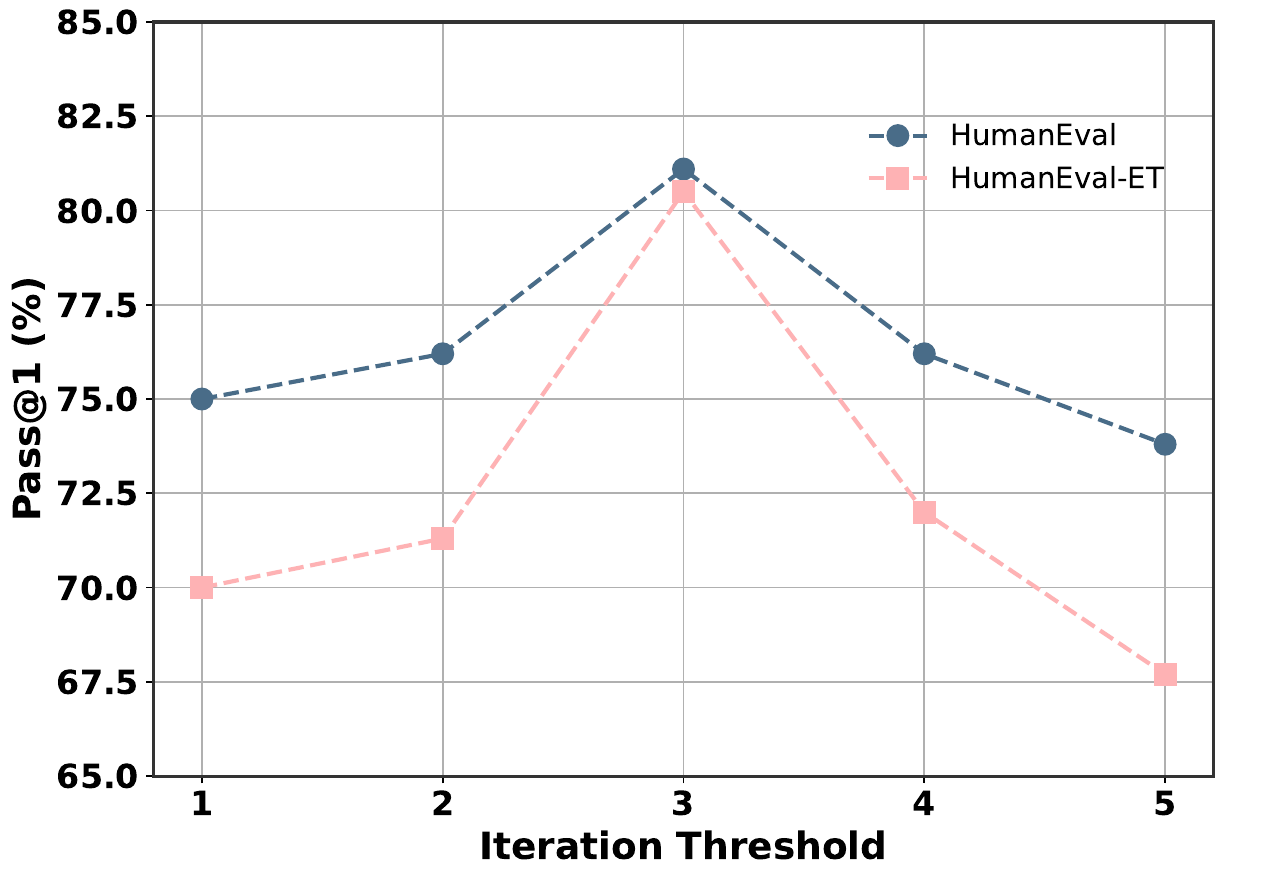}
    \vspace{-7px}
    \caption{Pass@1 results under different repair rounds on HumanEval datasets}
    \label{fig:pass1_results}
    \vspace{-7px}
\end{figure}

In CodeCoR, the number of repair rounds is a key factor for the performance of the framework. 
The coding repair process of each code snippet is stopped when the code cannot be repaired to pass more test cases. In this experiment, we force the code repair process to stop by setting different numbers of repair rounds. The results in Figure \ref{fig:pass1_results} show that the overall performance of the four agents reaches the best when the number of repair rounds is 3.

\subsection{Hyperparameter Analysis of Candidate Budget}

We analyze how the number of candidates generated by each agent affects CodeCoR on HumanEval.
As shown in Table~\ref{tab:candidate_budget}, cfg1--cfg4 gradually increase the numbers of prompt, test, code, and repair candidates.
The prompt, code, and repair candidates are increased together to reduce the number of variables.
The number of test candidates is slightly larger because generated tests are used not only for code generation, but also for scoring and pruning.

\begin{table}[htbp]
\centering
\caption{Hyperparameter analysis of candidate budget on HumanEval. 
P/T/C/R denote prompt/test/code/repair candidates.}
\vspace{-7px}
\label{tab:candidate_budget}
\setlength{\tabcolsep}{5pt}
\renewcommand{\arraystretch}{1.08}
\resizebox{0.75\columnwidth}{!}{
\begin{tabular}{lcccccc}
\toprule
\textbf{Config} 
& \textbf{P} 
& \textbf{T} 
& \textbf{C} 
& \textbf{R} 
& \textbf{Pass@1(\%)} 
& \textbf{Cost(\$)} \\
\midrule
cfg1 & 1 & 1 & 1 & 1 & 87.80 & 0.0020 \\
cfg2 & 2 & 3 & 2 & 2 & 88.41 & 0.0063 \\
cfg3 & 3 & 5 & 3 & 3 & \textbf{89.63} & 0.0110 \\
cfg4 & 5 & 7 & 5 & 5 & 80.49 & 0.0201 \\
\bottomrule
\end{tabular}
}
\vspace{-8px}
\end{table}

The results show that increasing the number of candidates is helpful only within a moderate range.
From cfg1 to cfg3, Pass@1 increases from 87.80\% to 89.63\%, suggesting that more candidates can provide better choices for pruning.
However, cfg4 drops to 80.49\% while having the highest cost.
This indicates that too many candidates may introduce more noisy intermediate outputs and make pruning harder.
Therefore, CodeCoR uses cfg3 as the recommended configuration for HumanEval-like tasks.

\subsection{Search Stability and Candidate Diversity Analysis}

We further analyze CodeCoR from two aspects: pruning/search stability and candidate diversity.
These analyses help clarify how CodeCoR's intermediate candidates behave across runs and whether multiple generated candidates provide meaningful alternatives for pruning.

\textbf{\textit{1) Pruning/search stability.}}
Table~\ref{tab:stability_diversity} shows that the retained intermediate candidates have low Jaccard overlap across three HumanEval runs.
This indicates that CodeCoR may follow different intermediate search trajectories under different random seeds.
However, the final pass/fail agreement remains high at 0.9350, suggesting that the final task-level outcomes are relatively stable even when the intermediate candidates vary.

\textbf{\textit{2) Candidate diversity.}}
Table~\ref{tab:stability_diversity} also shows that CodeCoR maintains non-trivial candidate diversity.
Prompt diversity stays around 0.86, test diversity around 0.55--0.57, and code diversity around 0.47--0.49.
This indicates that the generated candidates are not near-duplicates and can provide meaningful alternatives for pruning and selection.

\begin{table}[htbp]
\centering
\caption{Search stability and candidate diversity analysis of CodeCoR on HumanEval.}
\vspace{-7px}
\label{tab:stability_diversity}
\footnotesize
\setlength{\tabcolsep}{5pt}
\renewcommand{\arraystretch}{1.08}
\resizebox{0.7\columnwidth}{!}{
\begin{tabular}{lc}
\toprule
\textbf{Metric} & \textbf{Value} \\
\midrule
Avg. kept-plan Jaccard & 0.0000 \\
Avg. kept-test Jaccard & 0.0577 \\
Avg. kept-code Jaccard & 0.1420 \\
Final pass agreement & 0.9350 \\
\midrule
Prompt diversity & 0.8579--0.8593 \\
Test diversity & 0.5491--0.5658 \\
Code diversity & 0.4741--0.4933 \\
Unique assertion ratio & 0.6356--0.6435 \\
\bottomrule
\end{tabular}
}
\vspace{-8px}
\end{table}

Overall, CodeCoR's intermediate search process is sensitive to random seeds, but the final task-level outcomes remain relatively stable.
The diversity results further show that multiple generated candidates provide useful alternatives before pruning, supporting the design of candidate generation followed by selection.

%% file: Threats_to_Validity.tex
Our study demonstrates promising results. We have identified the following threats to validity:

We ensured a consistent experimental setup across all trials to minimize variations, although minor fluctuations in execution environments could still introduce variability. 
To reduce the likelihood of random factors influencing our conclusions, we conducted 10 rounds of experiments for each trial and averaged the results.
Despite these efforts, LLM-based generation and pruning may still introduce stochastic variation, especially when multiple candidates are sampled and evaluated.
Our ablation study validates pruning at the workflow level, but it does not fully prove that every individual pruning decision is always correct. 
In particular, we do not claim that every retained candidate is necessarily better than every pruned candidate. 
A fine-grained retained-versus-pruned candidate analysis is left for future work.

We considered construct validity by using execution-based metrics to evaluate functional correctness. 
Specifically, Pass@1 is used as the primary metric because it measures whether the generated code passes the benchmark-provided test suites. 
This choice is aligned with the main goal of code generation evaluation, namely producing functionally correct programs rather than textually similar code.
CodeBLEU is reported only as a complementary metric to reflect structural and semantic similarity to reference implementations.

Although our experiments were designed to reflect practical software development scenarios, the specific datasets and settings used might still limit the broader applicability of our findings. We mitigate this limitation by extending our evaluation to full DevEval and full RAL-Bench, which cover repository-level and application-level code generation. Nevertheless, automated benchmarks cannot fully replace controlled user studies for measuring developer productivity, usability, trust, and cognitive load. To address this, we plan to validate our approach on a wider range of datasets, such as SWE-bench~\citep{jimenez2023swe}, and in different environments in future research. Developer-centered evaluation is left as future work.

%% file: Conclusion.tex
This paper proposes CodeCoR, a multi-agent framework designed to realize self-reflection in code generation through intermediate output checking and pruning. 
It features a comprehensive workflow that allows for the checking and evaluation of intermediate outputs during the generation process.  
By employing four agents 
along with various pruning methods for pruning low-quality outputs, CodeCoR reduces the propagation of low-quality outputs and improves code generation performance, achieving an average Pass@1 score of 77.13\% on four public datasets and outperforming representative prompting and multi-agent baselines under our evaluation setting. 
The experimental results demonstrate that the self-reflective capability of CodeCoR improves the accuracy of code generation while maintaining manageable system-level overhead. 
Our replication package, including source code and prompt examples, is available at \url{https://github.com/Wwstarry/CodeCoR}.